# Arabia: from craters to stone circles

**Amelia Carolina Sparavigna**
Dipartimento di Fisica, Politecnico di Torino
Corso Duca degli Abruzzi 24, Torino, Italy

**Abstract**: The Arabia Shield has a volcanic nature inside. A region of the Western Saudi Arabia is in fact covered with vast fields of lava known as harraat. These lands are spotted by many stone circles and other quite interesting archaeological remains of the Neolithic period, such as the "desert kites", the hunters used to guide the game across the harrah in some corrals. With Google Maps, we can observe both sceneries, the volcanic nature of the land and a portrait of Arabia during the Neolithic times.

**Keywords**: Satellite maps, Landforms, Artificial landforms, Image processing, Archaeology

Geologically, Arabia lies on a tectonic plate of its own, the Arabian Plate, which is moving away from Africa, creating the Red Sea, and crushing northward with the Eurasian plate. The Romans subdivided this peninsula in three regions. One was the Arabia Petraea covering a wide region comprising the southern Syria, Jordan, the Sinai Peninsula and the north-western Saudi Arabia. Arabia Petraea was a roman province, with Petra as capital. The other two regions were the Arabia Deserta, the desert interior of the peninsula and the Arabia Felix, corresponding to the modern Yemen, which enjoys a more wet climate [1-3].

The Arabian peninsula has a central plateau, the Nejd, sloping eastward from the mountains along the coast of the Red Sea, to the shallow waters of the Persian Gulf, with a crescent of sand and gravel deserts toward east. The huge deserts are the stony Nefud in the north, the Dahna and the sands of Rub' Al-Khali. Ranges of mountains run parallel to the Red Sea coast on the west and at the south-eastern end of the peninsula.

Most of the Arabian rivers are wadis, that is rivers which are dry except during the rainy season. The presence of several aquifers creates the oases, when the flow of water reaches the surface of the desert. Under the sand of deserts, in 1932, oil was found. It could then happen that Arabia is imagined with a stereotype of sand-and-gravel deserts, punctured by oases and oil-wells, neglecting a quite important aspect of the region. This is the volcanic nature of the Arabian Shield, the geological name for much of the western Arabian peninsula. The Western Arabia is not only covered with sand, it is also clad with vast fields of lava. In Arabic, these lava fields are known as harraat (singular, harrah; before a name, harrat). Harraat together are creating large alkali basalt regions, covering some several thousand of square kilometers [4].

In 1256 an eruption threatened the city of Medinah, covering previous lava flows. The lava field near Madinah, which is known as Harrat Rahat is due to a volcanic activity of two million of years which is still active. The most recent eruptions on the Arabian Peninsula occurred in 1937, on a harrah near the town of Dhamar, in the north of Yemen. Before that, in 1846 an eruption took place on a volcanic Red Sea island. By counting the number of eruptions that have occurred on the northern Harrat Rahat, volcanologists estimate that, during the past 4,500 years, there have been 13 major eruptions, one every 346 years, on average [4]. The Arabian peninsula is then subjected to hazards coming from volcanic eruptions and from much more frequent earthquakes.

A survey with Google Maps shows the extensive lava fields covering the peninsula. Satellite imagery reveals the extent of harraat, with different colors of magma extruded in past epochs. Jet-black indicates the most recent flows. White areas are also observable, revealing places covered by silt and salt, which are residues of seasonal lakes, created by the lava flows that blocked some wadi rivers.

In Arabia, the maar craters, the landforms created by explosive eruptions, are huge: an explosion,

generated by the mixture of basaltic magma with subterranean water, produced the maar crater of al-Wahbah on the western margins of Harrat Kashib. According to [4], it is the Harrat Khaybar which is having the most distinctive volcanic scenery in all of Arabia, with the white cones of Jabal Bayda' and Jabal Abyad.

As a result of their detailed harraat research on-site work, Camp and Roobol discovered a new continental rifting activity [5,6], corresponding to a long volcanic system, the Makkah–Madinah–Nufud line, created by an active mantle up-welling. According to these researches, there is a fracture with crustal rifting in Arabia, which is much more recent of the Red Sea rift.

Figure 1 shows some of the craters in Arabia: one of them is crossed by a road and has a town, As Sayl, inside. The images are obtained from Google Maps, enhanced with image processing methods, as discussed in Ref. 7 and 8. Several other volcanic cones can be observed. One cone is attracting the attention because of its almost perfect shape and a close valley full of sand, with a crescent moon shape (see Figure 2, low panel). In fact, zooming the images to observe more details of the cone and its closer area, we can see small circles with diameters ranging from 10 to 15 meters, with a mound of stones inside. Knowing that near Titicaca Lake, worship places were composed by cairns of stones inside stone enclosures [9], I searched other such structures near the region of the crescent-moon valley, in particular, those which are clearly man-made. Figure 2 shows the result of this search, limited to the area with high-resolution satellite imagery: the markers are denoting the positions of stone circles. Of course, Figure 2 is not claiming to be exhaustive: surely more places have to be marked in this map. The figure is just trying to determine a rough distribution of these places. Figure 3 shows in more detail the positions of stone circles on a landform which seems a peninsula on a lake of sand. The following Figure 4 gives some pictures of these stone circles.

After searching with Google for details and references about possible stone circles in Arabia, some interesting web pages have been found. These pages show the archaeological evidence of the Neolithic period in Arabia. Among the Neolithic structures there are stone circles, lines and "desert kites", which were kite-shaped Neolithic stone fences, probably used as animal traps. There is also abundant archaeological evidence of Neolithic communities over harraat of Rahat and Khaybar, where thousands of tumuli and stone fences, keyhole-shaped, kite-shaped and circular, cover extensive areas [4]. The absolute time period of Neolithic in Arabia was 11,000-5,750 before present. For the bulk of the peninsula the term Neolithic is used to cover the period from 8,000 to 5,750 before present [10].

One quite interesting web page is Ref.11: the author, pen-name KenGrok, is discussing his researches of the Neolithic stone structures, using the satellite imagery. He found stone circles and other figures all situated at the peaks of hills or at the edges of plateaus. In [11], an article is reported, published in 1977 by the Sydney Morning Herald, telling that "enigmatic circular stone formations, reminiscent of those found in Europe, are scattered throughout this arid country on hilltops and valleys remote from human habitation. The rings are formed by stone walls that are 30 to 60 centimeters tall and range in diameter from five meters to more than 100 meters. No legends cast light on their origins or purpose and theories are myriad. Amateur archaeologists have noted that many of the rings have "tails," one or more appendages that sometimes stretch out for hundreds of meters across the wilderness. ... The English-language daily newspaper Arab News has speculated that a cluster of stone rings 60 km north of the Red Sea port of Jedda may be ancient grave sites. The walls were too low to have served as sheep or goat pens, the newspaper reasoned. Mr. Ron Worl, of the US Geological Survey [12], ... concluded that the stone rings could be the desert equivalent of rock carvings, ancient signposts that point the way to freshwater springs or caravan routes. Several of the "tails" led to water or old desert paths" [11-13].

After his survey with satellite, the author of [11] noted that only basic shapes were used and that these structures are not distributed evenly or at random. They appear most often along what must have been travel routes, usually found at prominent points of sorts (see for instance, Figure 3).

In [11], the different shapes of these Neolithic structures are listed: stone circles, often perfectly round with a cairn at the centre, stone circles with triangles, triangles and mounds of two kinds, round with flat top or with a depression or hole in the middle. In fact, as observed in [11], some mounds may be small volcano vents or cones, because from satellite imagery only, it is hard to tell whether they are man-made or geological features. Other structures have the shape of needles, lines or tails. There are enclosures, sometimes with round structures or irregulars: according to [11], these enclosures were very old dwelling and/or livestock areas.

Arabia is also spotted by the so-called "desert kites". Soon after air travel began over the Arabian peninsula pilots reported these big structures, and perhaps, this is why these objects result as quite deeply studied [4,14,15]. Several area in Jordan, Saudi Arabia, Syria and Iraq are covered with these man-made prehistoric kites, that sometimes are spanning kilometers of desert. Some researchers date them to Neolithic periods [14,15], and agree on the fact that they were used as hunting traps, to herd game in the ending enclosure of the kite.

Desert kites are basically triangular shapes with one end opened. It consist of two converging more or less straight walls, that create a funnel which ends with a narrow passage to a sort of pen. Typically there are two, three or more circular enclosures on the edge of this corral. There are archaeological rock art images defining these structures as hunting traps [16] and depicting the true role of the 'walls' of kites, that lead to ending enclosure. These walls are low and then not able to stop any game. These walls are not walls at all: they are the basements, in rocky harraat, to stick some poles in the ground and build a fence with branches. Moreover, as told in [16], these hunting traps where not designed to just capture and then kill animals, but also for the conservation of food, keeping the game alive, in small huts at the end of the kites. "What we have here is evidently a first attempt of domestication of animals … We probably do not speak of real domestication, but a step between hunting and organized intentional raising of animals" [16]. Of course, some scholars do not agree with such conclusions [14].

Another interesting point is raised [16]. In the Khaybar area, there are remains of Neolithic villages very close to the hunting desert kites. Linking the desert kites with the remains of houses and villages could be then a mistake, because wild animals are avoiding places where people live. Khaybar area is full with ancient burial structures, sometimes placed inside the desert kites. A possible conclusion [16] is that these burial mounds are from later periods, quite after the creation of the desert kites that could be placed in the early Neolithic times, when people settled and started to domesticate animals. These conclusions on burial places and kites have been obtained just using satellite images. It seems then that satellites can make a portrait of a form of collective hunting in its early stages of evolution into animal domestication.

As we concluded in [9], the use of Google Maps or in general of satellite services devoted to the analysis of "landscapes and built environments", can give useful information for archaeological and historical studies, for those locations well-preserved and not destroyed by natural or human activities.

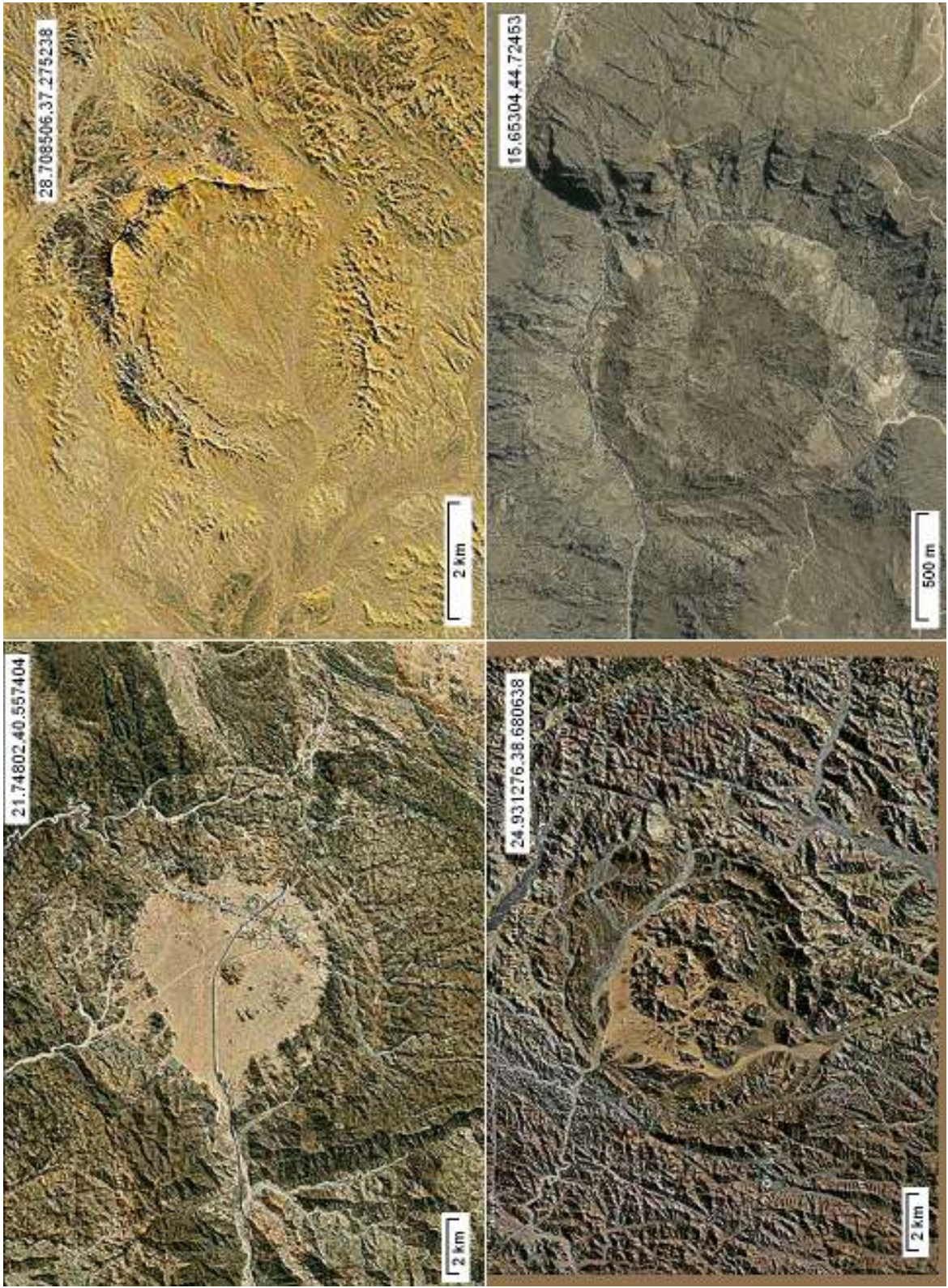

Figure 1: Some craters in Arabia. One of them has a road and a town, As Sayl, inside. Coordinates and scales are give in each panel. The images have been obtained after enhancing the Google Maps corresponding images.

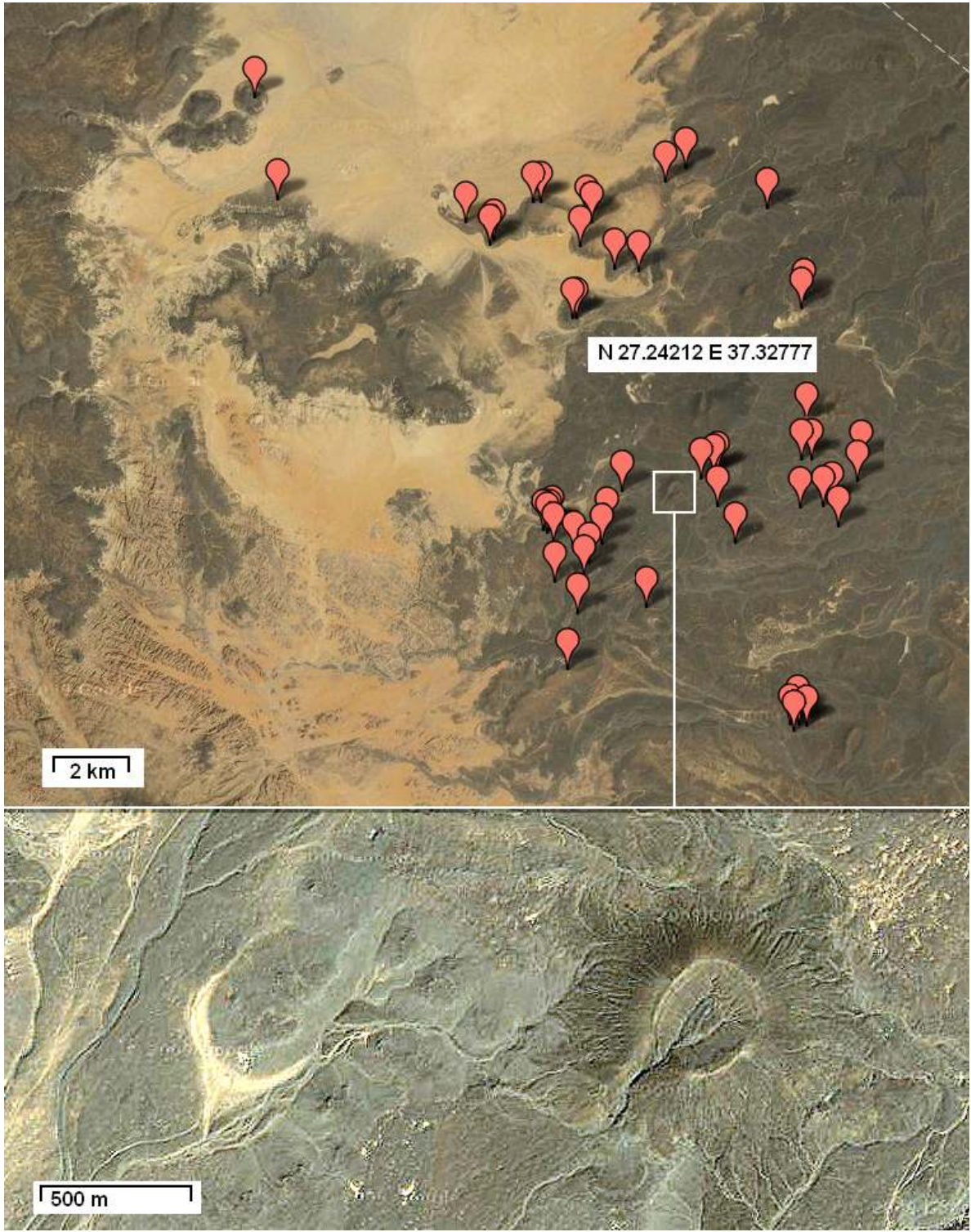

Figure 2: The lower panel shows an almost perfect volcanic cone and a near valley which seems a crescent moon. The upper panel shows the locations of stone circles and other stone enclosures. The position of the volcanic cone is inside the white square.

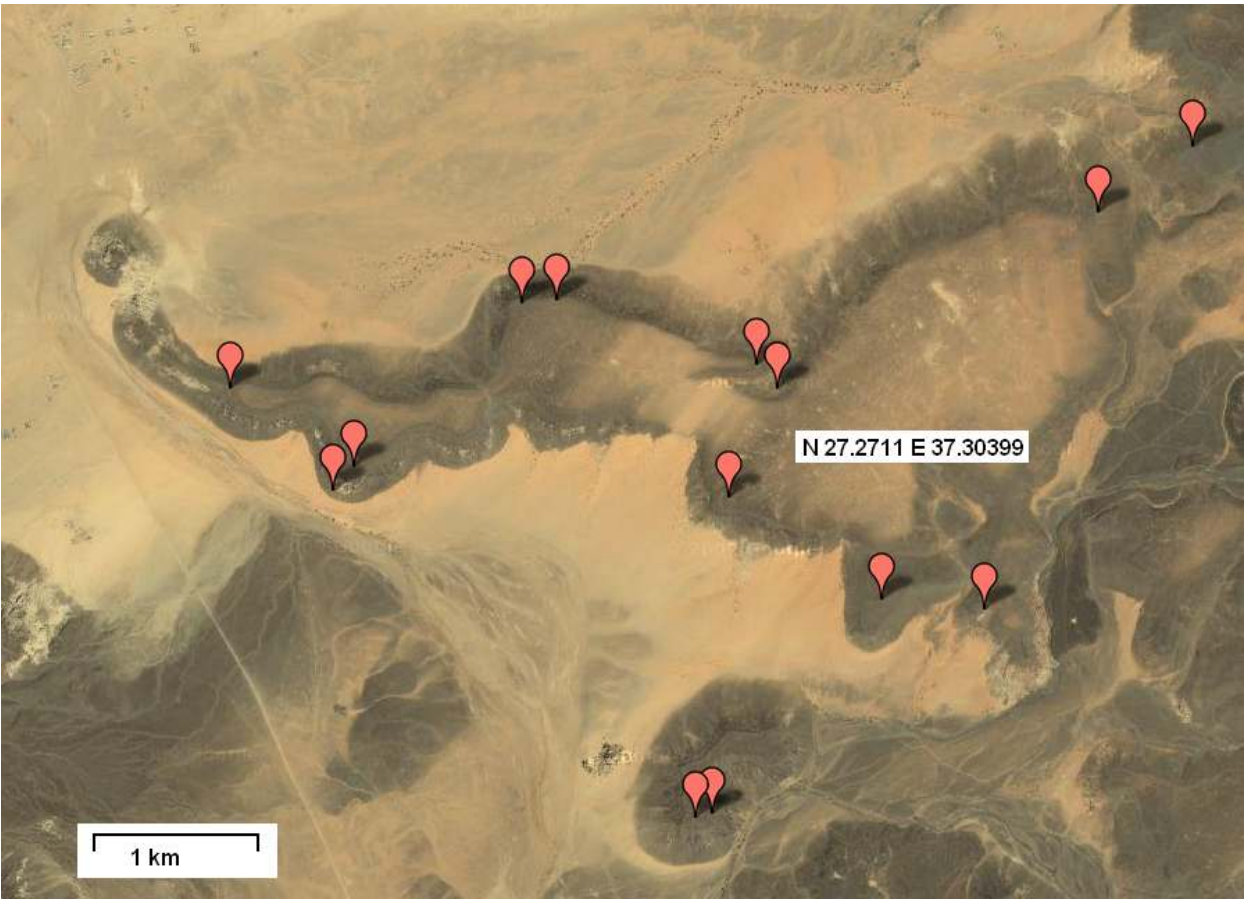

Figure 3: Positions of some stone circles.

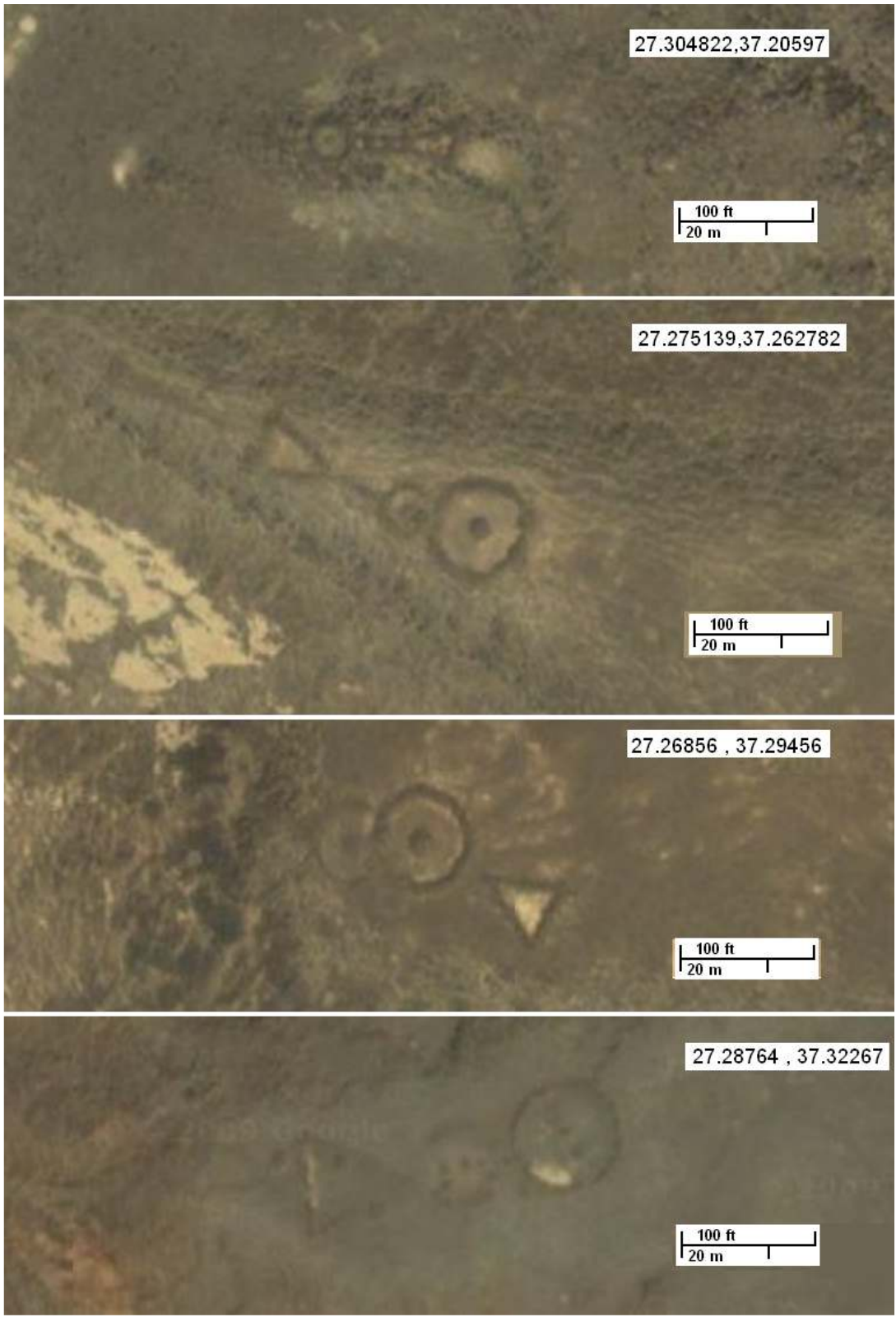

Figure 4: Some stone circles